\documentclass[11pt]{article}
\usepackage{a4,epsfig}
\newcommand{\xgo}{\mbox{$x_{\gamma}^{\rm obs}$}}
\renewcommand{\abstract}{}

\begin{document}

\begin{flushright}
IFT 2001/8 \\
\end{flushright}

\bigskip

\begin{center}
\section*{Heavy quark photoproduction at THERA}

{\bf P. Jankowski, M. Krawczyk}\\
Institute of Theoretical Physics, University of Warsaw\\
Warsaw, Poland\\
{\tt pjank@fuw.edu.pl, krawczyk@fuw.edu.pl}\\
{\bf M. Wing}\\
McGill University, Montreal\\
Canada\\
{\tt wing@mail.desy.de}

\end{center}

\begin{center}
\subsubsection*{Abstract}
\end{center}

\abstract{Measurements of heavy quark production at a possible future $ep$ facility, 
THERA, would provide valuable information on the structure of the photon. QCD 
cross-section predictions are made at LO for both charm and beauty production and 
their sensitivity to current parametrisations of the photon parton densities is 
investigated.}

\bigskip

\section{Introduction}

Heavy quark production can provide information on the structure of the real photon 
as well as provide stringent tests of QCD by itself. In this paper the potential of 
the possible future $ep$ collider, THERA, to give new information on the real photon 
structure through charm and beauty quark photoproduction is studied. The potential of 
the collider is also compared with other existing or planned colliders.

Photoproduction processes are dominated by the exchange of quasi-real photons 
(virtuality of the photon $Q^2\ll 1$GeV$^2$). Their energy distribution can be 
described by an effective (real) photon energy spectrum, i.e. using the 
Weizs\"acker-Williams approximation. A hard scale is provided by a large transverse 
momentum or the mass of the produced heavy quark $Q$. In this paper the region 
$p_T \geq m_Q$ is studied. In LO QCD, two processes contribute to the photoproduction 
of heavy quarks:  the direct process, in which the photon couples directly to a
parton in the proton, and the resolved process, in which the photon acts
as a source of partons, one of which scatters from a parton in the proton. Heavy 
quarks can be produced in the final state or they already exist in the initial state as 
partons in the photon and proton, depending on the scheme used in the calculations.

If not stated otherwise the computations were performed with the THERA electron and 
proton beams energies of 250 GeV and 920 GeV, respectively. In the first section 
results are presented on LO QCD calculations of inclusive charm quark production, 
$ep\to c/\bar{c}\: X$, at the THERA collider. In particular the possible information 
this process can reveal on the structure of the photon is discussed. The second section 
uses MC expectations to investigate events where at least one heavy quark and two jets 
are produced. Further requirements on a final state muon from a semi-leptonic decay 
are also imposed and sensitivity to current parametrisations of the photon investigated.

\section{Leading order QCD predictions of charm quark production}

LO calculations were performed in two schemes which differ by the number of quark flavours 
which are considered to be part of the structure of the photon and proton, and can thus 
take part in the process as partons. The massive scheme, the so-called Fixed Flavour 
Number Scheme (FFNS), assumes that both the photon and proton consist of gluons and 
only light ($u$, $d$ and $s$) quarks whilst heavy quarks are produced in the final 
state. The massless Variable Flavour Number Scheme (VFNS), considers heavy 
quarks as active flavours. Therefore such partonic processes, $Q+g\to Q+g$ or 
$Q+\bar{Q}\to Q+\bar{Q}$, with $Q$ denoting a heavy quark, are also allowed. 
(Events in which the heavy quark is one of the active partons are also referred
 to as ``flavour-excitation''.) In the FFNS scheme the mass of the produced charm quark is always 
kept nonzero in the calculations while in the VFNS all flavours are treated as massless.

In the massive (FFNS) calculations the number of active flavours ($N_f$) is taken to be 
3. In the massless (VFNS) scheme it varies from 3 to 4 depending on the value of the 
hard (factorisation and renormalisation) scale, $\mu$. The charm quark is included in 
the computation provided that $\mu>m_c$ with the mass of the charm quark set to 
$m_c=1.5$GeV. Also the QCD energy scale, $\Lambda_{QCD}$, which appears in the one loop 
formula for the strong coupling constant $\alpha_s$ is affected by the change of the 
number of active flavours, so is denoted as $\Lambda_{QCD}^{N_f}$. The 
scale is taken to be $\Lambda_{QCD}^{3} = 232$ MeV and $\Lambda_{QCD}^{4} = 200$ MeV as 
in the GRV\footnote{The old GRV('92) parametrisation was used in this analysis. It treats 
heavy quarks as massless 
above the heavy quark threshold region ($W\gg 2m_Q$), and moreover latest experimental results
\cite{GRVH1} indicate that its gluon distribution is closest to the data.} parametrisation 
of the photon~\cite{GRV}. If not stated otherwise $\mu$ is taken to be the transverse mass of the 
produced charm quark, $m_T=\sqrt{m_c^2+p_T^2}$, and the differential cross-section, 
$d^2\sigma/d\eta dp_T^2$, with $p_T=10$ GeV is calculated. This means that $\mu \sim p_T$ 
and therefore $N_f=4$ in case of the VFNS calculations. The spectrum of the 
quasi-real photons taking part in the process is described by the Weizs\"acker-Williams 
(EPA) formula,
\begin{equation}
f_{\gamma/e}(y) = \frac{\alpha}{2\pi}\left[ 
\frac{1+(1-y)^2}{y}\ln \frac{Q^2_{max}}{Q^2_{min}} 
+ 2m_e^2y\left(\frac{1}{Q^2_{max}} - \frac{1}{Q^2_{min}}\right)  \right],
\end{equation}
where $Q^2_{max}=4$GeV$^2$, $Q^2_{min}=m_e^2y^2/(1-y)$ and $y$ is the fraction of
the electron's four-momentum carried by the photon ($y=E_{\gamma}/E_e$). No requirements 
were placed on $y$. Finally the LO CTEQ5L~\cite{CTEQ} and GRV~\cite{GRV} were used as the 
parton parametrisations of the proton and photon, respectively.

\bigskip

To show the power of the THERA collider for heavy quark physics it is
useful to compare the cross-section for charm production that can be achieved in
THERA with cross-sections expected for other existing or planned 
accelerators. In Fig.~\ref{akcel}, a comparison with results from the $ep$ HERA 
($E_e= 27.5$ GeV, $E_p=920$ GeV), LEP ($E_e=90$ GeV), a Linear Collider (LC) 
and a Photon Collider (PC) both with $E_e=250$ GeV are presented for 
two options of THERA, $ep$ and $\gamma p$. Real photons in the PC and 
$\gamma p$ THERA option can be achieved through the Compton backscattering laser 
photons from the electron beam \cite{Ginz,Tel} (the full theoretical spectrum for 
$0<y<0.83$ was used). It can be seen that THERA provides the largest cross-sections 
over the whole range in $\eta$ and has the potential to probe regions kinematically 
inaccessible at other colliders.

\bigskip

\begin{figure}[htp]
\begin{center}
{\input{akcelerg1.tex}}
\end{center}
\caption{Inclusive charm quark production in processes including real
photon interactions calculated in the VFNS scheme for $p_T=10$ GeV for five
accelerators: THERA ($ep$ and $\gamma p$ options) and HERA 
($E_e=27.5$ GeV, $E_p=920$ GeV) ($ep\to c/\bar{c}\: X$)
, LEP ($E_e=90$ GeV) \& LC ($E_e=250$ GeV) ($e^+e^-\to e^+e^- \: c/\bar{c}\: X$) with 
quasi-real WW photons and PC 
($e^+e^-\to e^+e^- \: c/\bar{c}\: X$) with real photons achieved through the 
Compton backscattering laser photons from the electron beam and $E_e=250$ GeV 
\cite{Ginz,Tel}). CTEQ5L proton and GRV LO photon parton parametrisations were 
used.}
\label{akcel}
\end{figure}

It has been demonstrated elsewhere~\cite{KKKS}, that there is a large dependence of the 
cross-section for heavy quark production on the choice of scheme for the case of HERA 
energies. It is also true for the THERA collider. Direct and resolved contributions to the cross-section for charm production obtained
using the two discussed schemes clearly show this dependency, as displayed 
in Fig.~\ref{sch1}. Note that the resolved and direct cross-sections are similar for the 
VFNS scheme whereas in the FFNS scheme, the resolved cross-section is small compared 
to that of the direct contribution. The small difference between the direct cross-sections for the two 
schemes is due to the mass term in the matrix elements and phase space integral. 
Figure~\ref{sch2} presents the cross-sections with their dependence 
on the choice of the $\mu$ varied from $1/2m_T$ to $2m_T$. The dependence on the choice 
of scheme is most significant for resolved processes, varying by up to several orders of 
magnitude in the cross-section. In the case of direct processes the variation is relatively 
small and the subsequent total cross-section prediction variation is therefore dominated 
by the difference between the two schemes for the resolved process. The variation in the 
choice of scale leads to small effects in the cross-section in comparison with the scheme 
dependence as it only influences the value of $\alpha_s$ and the parton densities and not 
the set of allowed partonic processes.

\begin{figure}[htp]
\unitlength=1mm
\begin{picture}(0,0)(100,100)
\put(166,164){\bf \Large{(a)}}
\put(245,164){\bf \Large{(b)}}
\end{picture}
\centerline{\input{direct.tex} \input{resolved.tex}}
\caption{THERA. (a) Direct and (b) resolved contributions to the differential cross-section
of $ep\to c/\bar{c}\: X$ production calculated for $p_T=10$ GeV in the VFNS 
(solid lines) and FFNS (dashed lines) schemes using the CTEQ5L and 
GRV LO parton parametrisations of the proton and photon, respectively.}
\label{sch1}
\end{figure}

\begin{figure}[htp]
\centerline{\input{total.tex}}
\caption{THERA. Differential cross-section of $ep\to c/\bar{c}\: X$ production calculated 
for $p_T=10$ GeV in the VFNS (solid lines) and FFNS (dashed lines) schemes with three 
different values of the hard scale; $\mu = 1/2,1$ and $2 \times m_T$, using the CTEQ5L 
and GRV LO parton parametrisations of the proton and photon, respectively.}
\label{sch2}
\end{figure}

To test the sensitivity of inclusive charm production to the form of the parton densities 
in the photon, cross-sections for this process were computed with three different LO parton 
parametrisations of the photon: GRV~\cite{GRV}, GS96~\cite{GS} and SaS1D~\cite{SAS}. 
Sensitivity to the form of the current parametrisations of the photon structure function 
at the THERA collider is larger than at HERA as shown in Fig.~\ref{sens}a (the results of 
the VFNS calculation). Note that the SaS1D and GRV parametrisations give 
similar results. Figure~\ref{sens}b shows the uncertainty resulting from the 
variation of the hard scale by a factor of two compared to the central value. Here 
it is demonstrated that the differences between the cross-sections obtained using 
three parton parametrisations of the photon are significant for GRV (or SaS1D) and GS96 
in the region $-2.5<\eta<0$, despite this large theoretical uncertainty.

The differences between two current parton parametrisations of the 
photon, GRV and SaS1D, have previously been compared for LC and PC colliders~\cite{JKR}. 
More specifically, the relative difference $\frac{GRV-SaS1D}{GRV}$ of the cross-section, 
$d^2\sigma/d\eta dp_T^2$($ep\to c/\bar{c}\: X$), was calculated in 
both VFNS and FFNS schemes. Values of $\sim 0.1 - 0.4$ were found with the PC having the 
larger value. Similar numbers (up to 0.3) to those for the LC and PC could be achieved 
at THERA in the forward $\eta$ region. The massive FFNS gives the same results for the 
considered sensitivity as those of the massless VFNS scheme.

\begin{figure}[htp]
\centerline{\input{epcvf11ht.tex} \input{epcvf011con.tex}}
\caption{Differential cross-section for 
$ep\to c/\bar{c}\: X$ for $p_T=10$ GeV calculated
with the CTEQ5L parton parametrisation of the proton and three various LO parton 
parametrisations of the photon in the 
massless (VFNS) scheme. 
A) Lower lines correspond to the HERA collider energies (27.5$\times$920 GeV),
while upper to the THERA collider energies.
B) Lower and upper lines for each photon parametrisation show respectively the
lowest and greatest values of the cross-section calculated with different
values of the hard scale $\mu = 1/2,1,\; \& \; 2 \times m_T$ in the GRV case and 
$\mu = 1,2 \times m_T$ in the GS96 case.}
\label{sens}
\end{figure}

Sensitivity to the photon parton densities can be also tested through the ratio
of the resolved to the direct photon contributions. This is shown in Fig.~\ref{fig:ratio} 
for $d^2\sigma/d\eta dp_T^2$ in comparison with the results for the 
HERA collider. Again the results for THERA show a larger resolved cross-section 
and increased sensitivity to the form of the current parametrisations. The sensitivity is also 
large in the central $\eta$ region where there is, experimentally, a large acceptance.

\begin{figure}[htp]
\centerline{\input{epcvf211dr.tex} \input{epcvf011dr.tex}}
\caption{The ratio of the resolved to the direct photon contribution of
$e^+e^- \rightarrow e^+e^- \: c/\bar{c}\: X$ (for $d^2\sigma/d\eta dp_T^2$)
for $p_T$=10 GeV calculated in LO with the CTEQ5L parton parametrisation of the proton and 
three various LO parton parametrisations of the photon in the massless (VFNS) scheme. 
A) HERA, B) THERA collider.}
\label{fig:ratio}
\end{figure}

\section{MC predictions of dijet production with charm and beauty quarks}

The dijet system can be formed from the production of a heavy quark pair such as in 
the boson-gluon fusion process. When the heavy quark is an active parton in the 
photon then the dijet system is formed from one heavy quark and a gluon. The other 
heavy quark in the event is produced at low transverse energy and forms part of the 
photon remnant.

A natural variable to describe dijet production is the observable \xgo, which is 
the fraction of the 
photon's energy producing the two jets of highest transverse energy~\cite{xgo};
\begin{equation}
\xgo = \frac{E_T^{\rm jet1}{\rm e}^{- \eta ^{\rm jet1}} 
           + E_T^{\rm jet2}{\rm e}^{- \eta ^{\rm jet2}}}{2yE_e},
\label{eq:xgamma}
\end{equation}
where $\eta^{\rm jet}$ is the pseudorapidity of the jet and $E_e$ is the electron energy 
and $y$ the fraction of the electron's energy carried by the photon.

\begin{figure}[htp]
\unitlength=1mm
\begin{picture}(0,0)(100,100)
\put(161,87){\bf \Large{(a)}}
\put(240,87){\bf \Large{(b)}}
\end{picture}
\begin{center}
~\epsfig{file=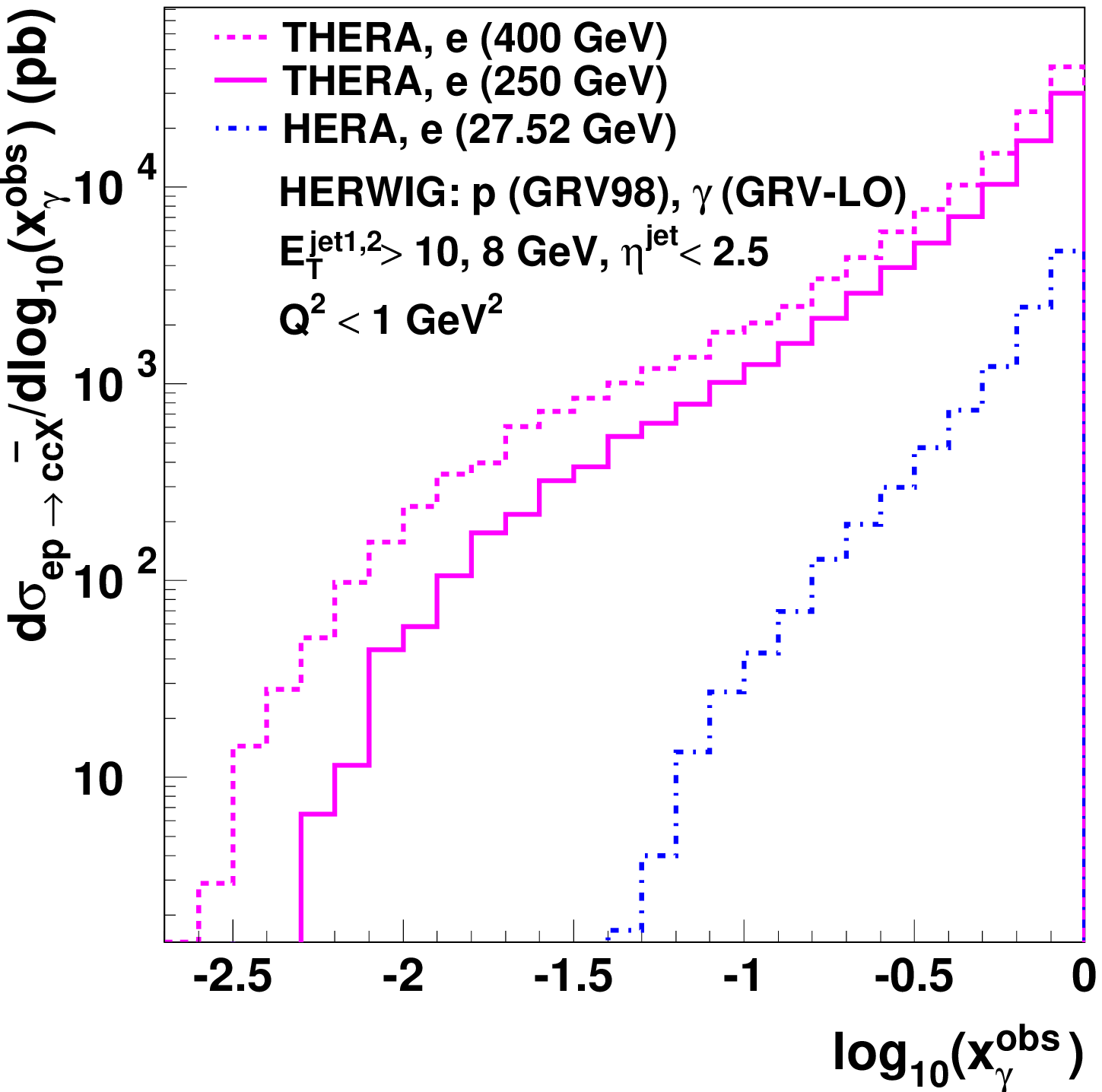,height=7.7cm}
~\epsfig{file=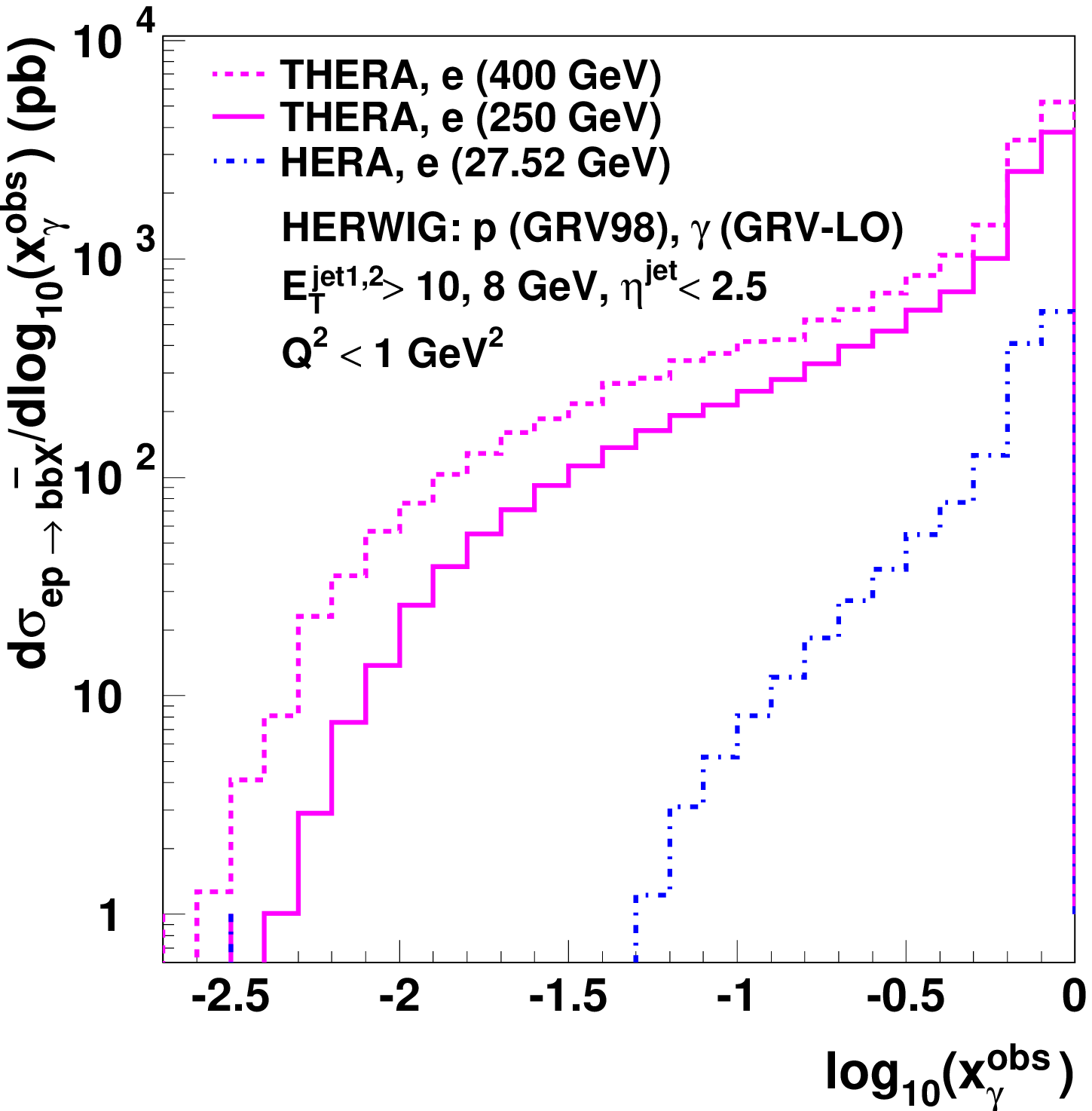,height=7.7cm}
\end{center}
\caption{The differential cross-section, $d\sigma/d{\rm log_{10}}\xgo$ for (a) charm and 
(b) beauty in dijet photoproduction at HERA and THERA as predicted by the {\sc Herwig} MC. 
The dashed and solid lines show THERA with an electron energy of 400 GeV and 250 GeV 
respectively and the dot-dashed lines indicate the expectation from HERA. The kinematic 
range is the same in all three cases, namely, $Q^2<1~{\rm GeV}^2$, with two jets, 
$E_T^{\rm jet1,2}>10,8$~GeV and $\eta^{\rm jet}<2.5$.}
\label{c_b_thera}
\end{figure}

Due to the increase in the beam energy, there is a corresponding order of magnitude 
extension in the minimum \xgo. This is shown in Fig.~\ref{c_b_thera} in which two THERA 
scenarios (with the electron beam energy $E_e=250$ GeV and $E_e=400$ GeV) are compared 
with HERA. A greatly increased cross-section possible at THERA over that at HERA for 
low$-$\xgo, particularly in beauty production, is seen. Realistic scenarios for the cuts 
on the jet quantities have been made of transverse energy $E_T^{\rm jet1,2}>$ 10,8 GeV 
and pseudorapidity $\eta^{\rm jet}<$ 2.5. The distribution is peaked at high-\xgo, 
consistent with direct photon processes, but has a significant cross-section at 
low-\xgo \ arising from resolved photon events.
 
\begin{figure}[htp]
\unitlength=1mm
\begin{picture}(0,0)(100,100)
\put(161,89){\bf \Large{(a)}}
\put(240,89){\bf \Large{(b)}}
\end{picture}
\begin{center}
~\epsfig{file=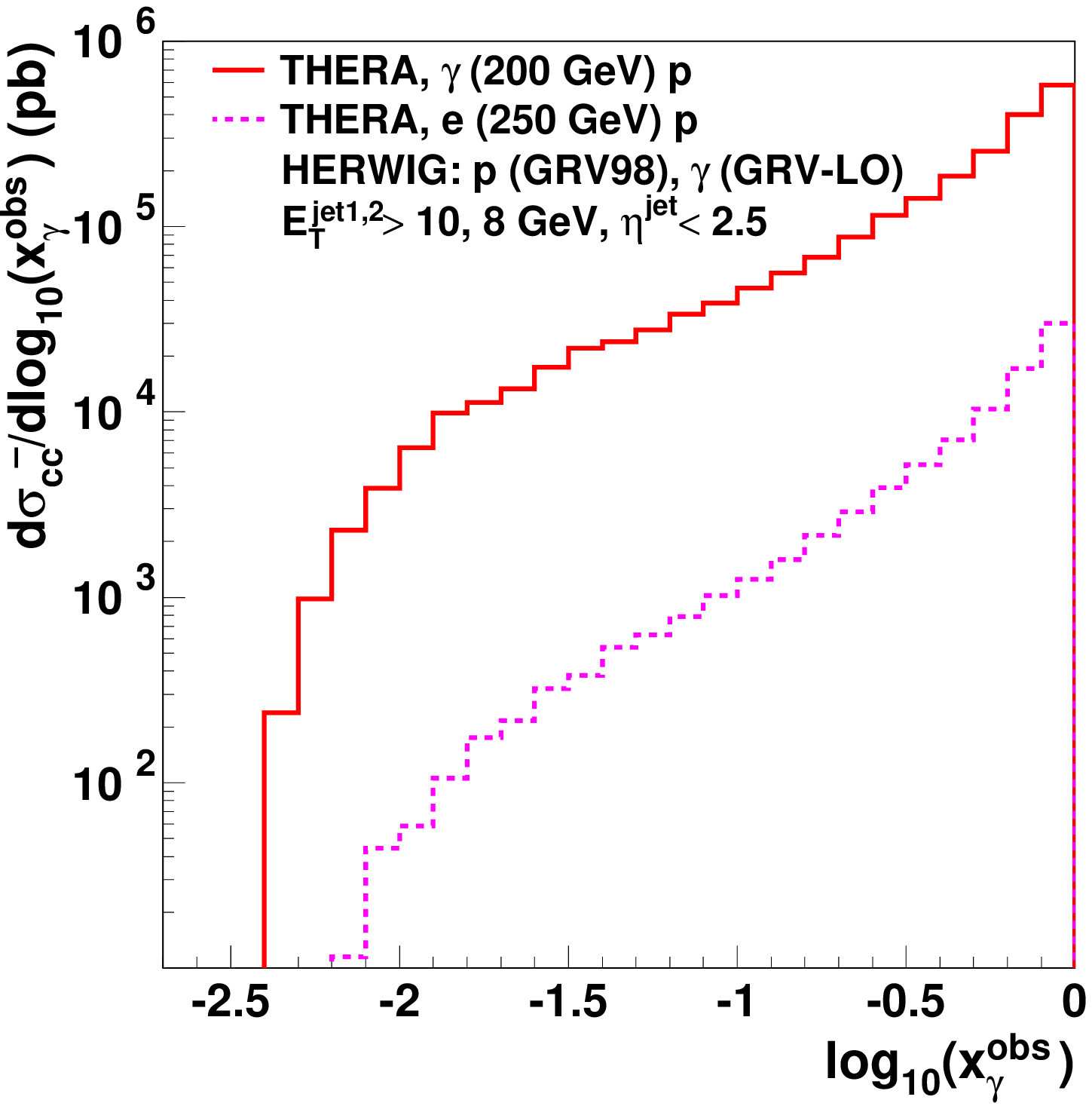,height=7.7cm}
~\epsfig{file=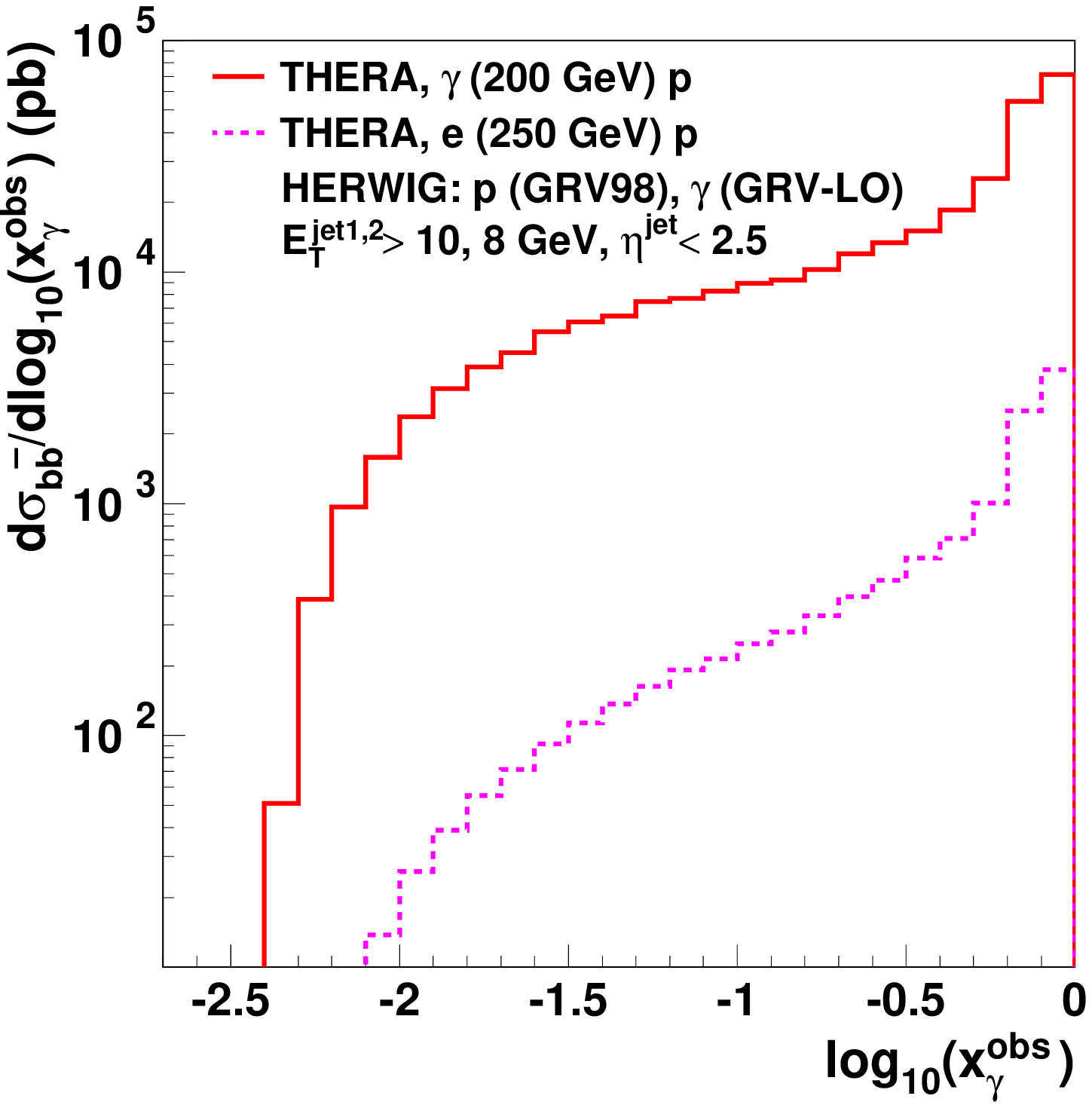,height=7.7cm}
\end{center}
\caption{The differential cross-section, $d\sigma/d{\rm log_{10}}\xgo$ for (a) charm and 
(b) beauty in dijet photoproduction at for $\gamma p$ and $ep$ options at THERA as predicted 
by the {\sc Herwig} MC. The solid lines show THERA with an electron energy of 250 GeV 
and the dashed lines the expectation with an photon energy of 200 GeV. The 
kinematic range is the same in all three cases, namely, two jets, 
$E_T^{\rm jet1,2}>10,8$~GeV and $\eta^{\rm jet}<2.5$.}
\label{c_b_gammap}
\end{figure}

Charm and beauty production at a $\gamma p$ collider has also been considered and 
compared with the nominal $ep$ scenario for THERA. The photon beam energy is 
assumed to be 0.8$E_e$, i.e. 200 GeV assuming an electron beam of energy 250 GeV. 
The predictions for the $ep$ and $\gamma p$ options with the same jet cuts are shown in 
Fig.~\ref{c_b_gammap}. The cross-section prediction for the $\gamma p$ option 
is a factor of $\sim$ 25 higher than the nominal $ep$ option. This factor arises simply 
from the Weizs\"{a}cker-Williams flux associated with the electron.

\begin{figure}[htp]
\unitlength=1mm
\begin{picture}(0,0)(100,100)
\put(138,87){\bf \Large{(a)}}
\put(213,87){\bf \Large{(b)}}
\end{picture}
\begin{center}
~\epsfig{file=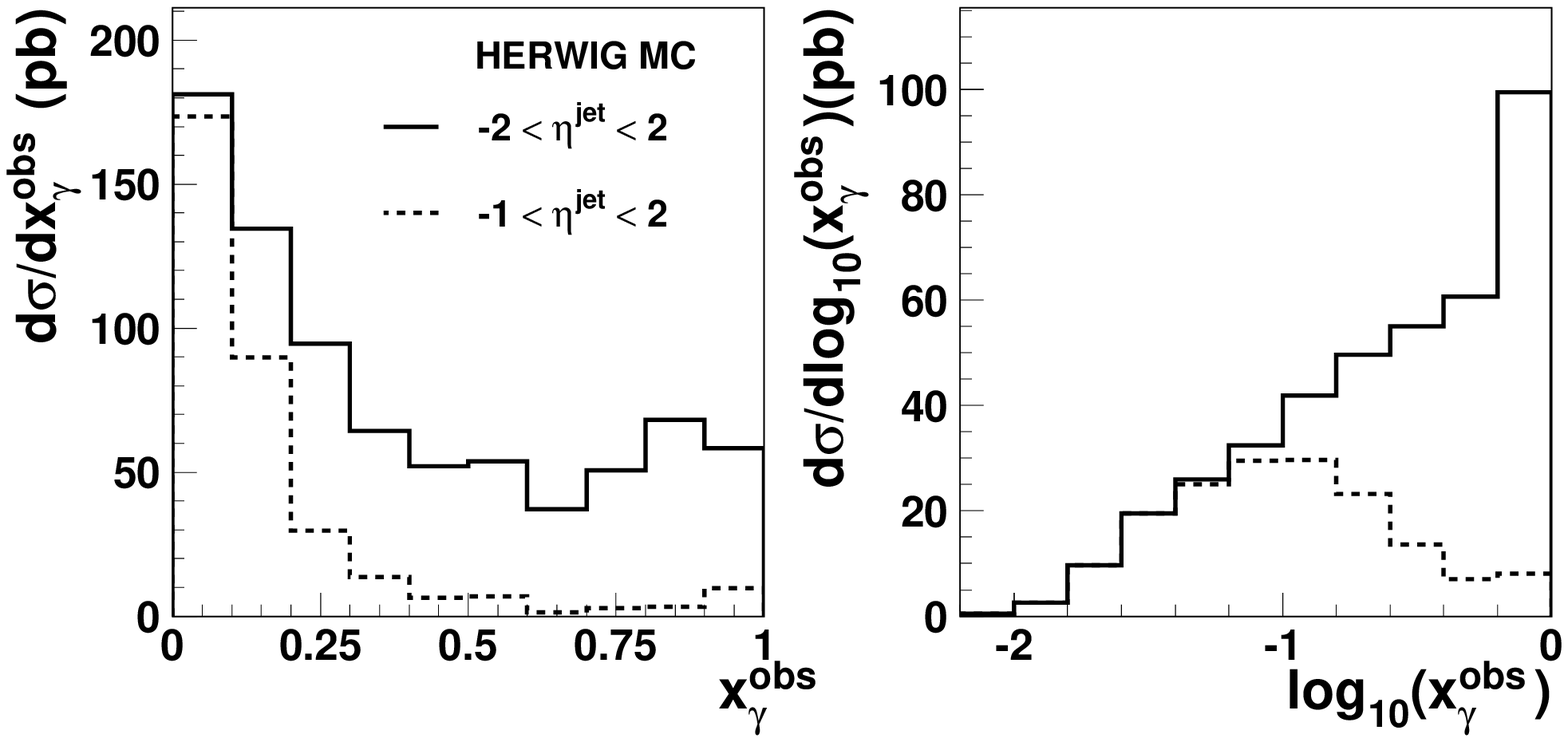,height=7.7cm}
\end{center}
\caption{The differential cross-sections, (a) $d\sigma/d\xgo$ and 
(b) $d\sigma/d{\rm log_{10}}\xgo$ for beauty in dijet photoproduction at THERA as 
predicted by the {\sc Herwig} MC. The events are selected in the range, $Q^2<1~{\rm GeV^2}$, 
$0.2<y<0.85$ with two jets of transverse energy, $E_T^{\rm jet1,2}>7,6$ GeV. The points 
represent when the jets are restricted to $-1<\eta^{\rm jet}<2$ and the solid lines when the 
jets are restricted to $-2<\eta^{\rm jet}<2$. A muon in the final state has also been 
required with transverse momentum, $p_T^\mu>2$ GeV and $|\eta^\mu|<2$.}
\label{eta}
\end{figure}

Further requirements can be made such that the kinematic range closely corresponds 
to what could be measured. Therefore the observation of a final state muon from the 
semi-leptonic decay of a beauty quark is imposed. The momentum and angular requirements  
imposed on the muon are typical of those at HERA. Considering jets of slightly lower 
transverse energy ($E_T^{\rm jet1,2}>$ 7,6 GeV), the effect on the low-\xgo \ 
component of pseudorapidity cut is shown in Fig.~\ref{eta}. It can be seen that 
for a pseudorapidity cut of $\eta^{\rm jet}>-1$, the direct peak at high-\xgo \ is 
drastically reduced with respect to a cut of $\eta^{\rm jet}>-2$. This demonstrates 
that by applying a cut of $\eta^{\rm jet}>-1$ interactions of almost exclusively the 
resolved photon are being probed.

\begin{figure}[htp]
\unitlength=1mm
\begin{picture}(0,0)(100,100)
\put(138,87){\bf \Large{(a)}}
\put(213,87){\bf \Large{(b)}}
\put(138,10){\bf \Large{(c)}}
\put(213,10){\bf \Large{(d)}}
\end{picture}
\begin{center}
~\epsfig{file=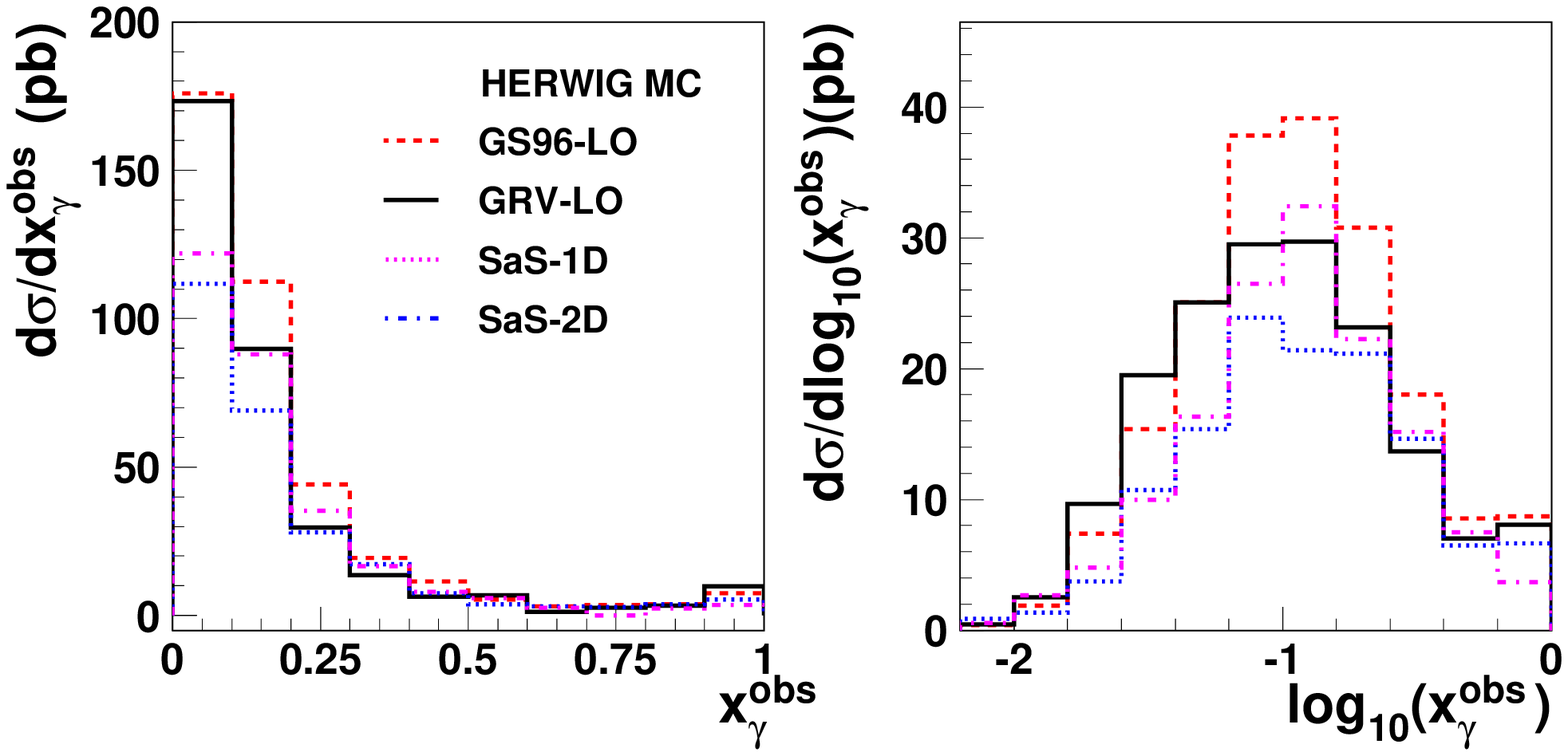,height=7.7cm}
~\epsfig{file=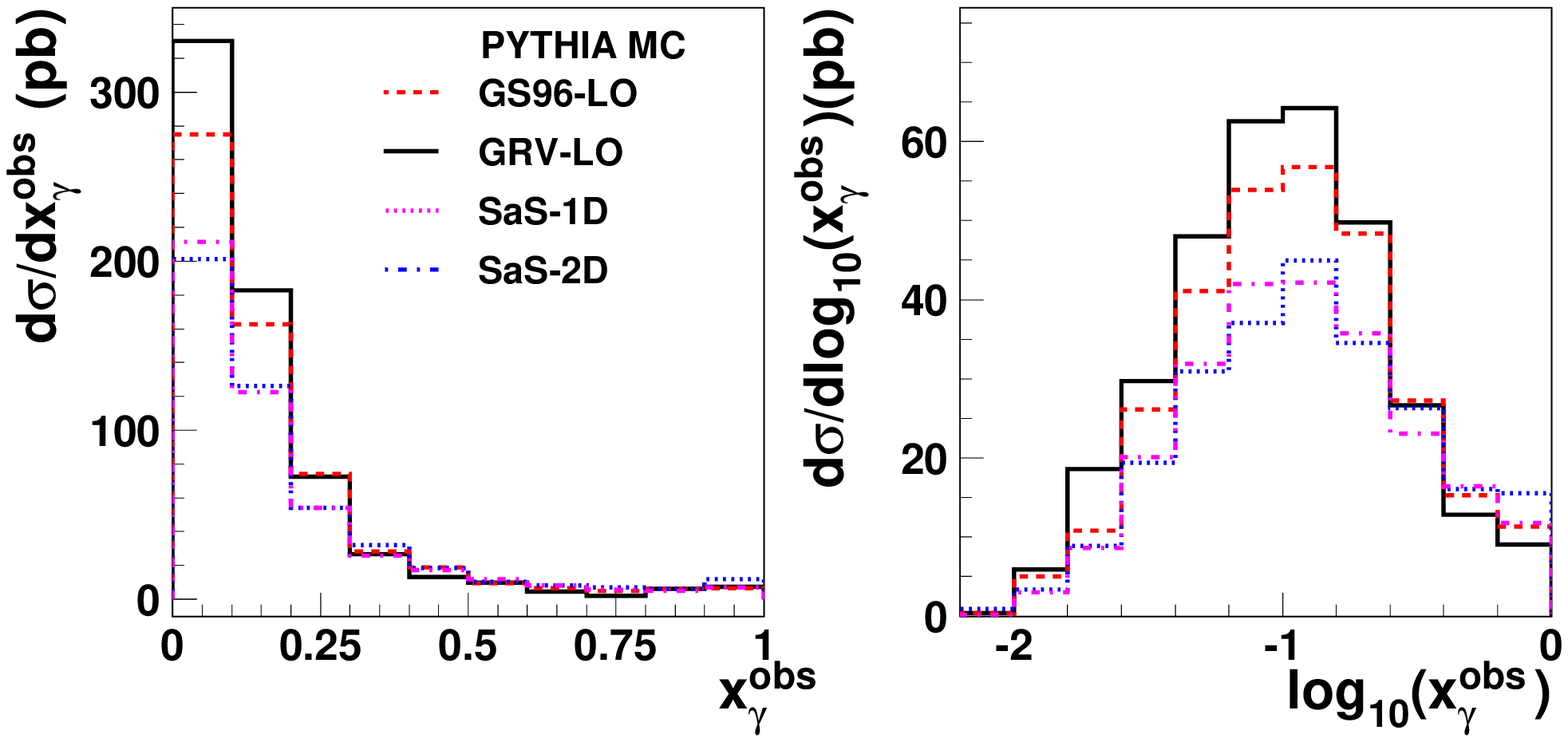,height=7.7cm}
\end{center}
\caption{The differential cross-sections, (a) $d\sigma/d\xgo$ and 
(b) $d\sigma/d{\rm log_{10}}\xgo$ for beauty in dijet photoproduction at THERA as 
predicted by the {\sc Herwig} MC. The differential cross-sections, (c) $d\sigma/d\xgo$ and 
(d) $d\sigma/d{\rm log_{10}}\xgo$ for beauty in dijet photoproduction at THERA as 
predicted by the {\sc Pythia} MC. Predictions for four different PDFs are shown; 
GS96-LO (open circles), GRV-LO (solid circles), SaS-1D (stars) and SaS-2D (diamonds).}
\label{photon_sf}
\end{figure}

This rather tight restriction on the pseudorapidity of the jets of $-1<\eta^{\rm jet}<2$ 
is imposed and the sensitivity of the cross-section to the current parametrisations 
of the photon structure function investigated for THERA. In Fig.~\ref{photon_sf} 
predictions are shown for events where a beauty quark decays into a muon. Predictions 
are shown using both {\sc Herwig}~\cite{herwig} and {\sc Pythia}~\cite{pythia} MC 
generators for four different LO photon PDFs. At 
values of \xgo=0.1, differences between the four parametrisations of up to $40\%$ are 
seen. The predictions from GRV and GS96 generally have the largest cross-sections at 
low-\xgo \ and also rise more quickly. The large difference in the {\sc Pythia} and 
{\sc Herwig} MC predictions comes mainly from the different default treatments of 
$\alpha_s$, different scales and hadronisation and the use of massive matrix elements in 
{\sc Herwig} and massless in {\sc Pythia} for the generation 
of flavour-excitation processes. 

\section{Summary}

It has been shown that the photoproduction of heavy quarks at THERA has a much increased 
cross-section relative to HERA. With the expected luminosity of about 100 pb$^{-1}$/year 
stringent tests of pQCD could be performed. It has also been shown that the cross-sections 
are sensitive to the structure of the photon, with the current parametrisations differing 
by significant amounts. The THERA collider also compares favourably with other current 
and planned machines studying heavy quark production, with the $\gamma p$ mode for 
THERA providing the largest cross-section.

\section*{Acknowledgements}

Some of use (PJ and MK) have been partly supported by the Polish State Committee
for Scientific Research (grant 2P03B05119, 2000-2001). MK is also grateful to the 
European Commission 50th Framework Contract HPRN-CT-2000-00149 and DESY for financial 
support.

\end{document}